\def\bd{\begin{displaymath}}\def\ed{\end{displaymath}}
\def\be{\begin{equation}}\def\ee{\end{equation}}
\def\bea{\begin{eqnarray}}\def\eea{\end{eqnarray}}
\def\nn{\nonumber}\def\lb{\label}

\def\a{\alpha}\def\c{\chi}\def\d{\delta}\def\e{\epsilon}
\def\f{\phi}\def\g{\gamma}\def\h{\theta}
\def\k{\kappa}\def\l{\lambda}\def\m{\mu}\def\n{\nu}\def\o{\omega}\def\t{\tau}
\def\y{\eta}\def\x{\xi}\def\z{\zeta}

\def\D{\Delta}\def\F{\Phi}\def\G{\Gamma}

\def\de{\partial}\def\per{\times}
\def\id{\equiv}\def\mo{{-1}}\def\ha{{1\over 2}}

\def\we{\wedge}\def\di{{\rm d}}\def\Di{{\rm D}}

\def\const{{\rm const}}
\def\arcsh{{\rm arcsinh}}\def\arcch{{\rm arccosh}}

\def\fe{field equations }\def\bh{black hole }

\def\tran{transformation }\def\coo{coordinates }

\def\gt{gauge transformation }
\def\cco{cosmological constant }

\def\gct{general coordinate transformations }\def\gts{gauge transformations }
\def\pb{Poisson brackets }

\def\ads{anti-de Sitter }
\def\poi{Poincar\'e }

\def\dpa{deformed \poi algebra }\def\psm{Poisson sigma model }
\def\td{two-dimensional }\def\trd{three-dimensional }

\def\PL#1{Phys.\ Lett.\ {\bf#1}}
\def\PRL#1{Phys.\ Rev.\ Lett.\ {\bf#1}}
\def\PR#1{Phys.\ Rev.\ {\bf#1}}\def\CQG#1{Class.\ Quantum Grav.\ {\bf#1}}

\def\PTP#1{Prog.\ Theor.\ Phys.\ {\bf#1}}

 \def\IJMP#1{Int.\ J. Mod.\ Phys.\ {\bf #1}}
\def\MPL#1{Mod.\ Phys.\ Lett.\ {\bf #1}} 
\def\PRep#1{Phys.\ Rep.\ {\bf#1}}
\def\AoP#1{Ann.\ Phys.\ {\bf#1}}
\def\hep#1{{\tt hep-th/#1}}

\def\lm{{1\over\l}}\def\eabc{\e_{ab}^{\ \ c}}

\def\NN{N^\pm}\def\XX{X^\pm}\def\AA{A^\pm}\def\QQ{Q^\pm}
\def\yy{\eta^\pm}\def\hh{\theta^\pm}\def\ww{\psi^\pm}
\def\EE{E_\pm}\def\cN{$\cal N$ }
\def\man{${\cal M}^3$}\def\pik{\left(1-{P_0\over\k}\right)}

\def\btz{\sqrt{\l^2r^2-M+{J^2\over4r^2}}}
\def\un#1{{\underline{#1}}}\def\ov#1{{\overline{#1}}}

\documentclass[12pt]{article}
\begin{document}

\begin{titlepage}
\vspace{.3cm}
\begin{center}
\renewcommand{\thefootnote}{\fnsymbol{footnote}}
{\Large \bf Solutions of deformed three-dimensional gravity}
\vskip 15mm%27.mm
{\large \bf {S.~Mignemi\footnote{email: smignemi@unica.it}}}\\
\renewcommand{\thefootnote}{\arabic{footnote}}
\setcounter{footnote}{0}
\vskip 7mm%1cm
{\small
Dipartimento di Matematica, Universit\`a di Cagliari,\\
Viale Merello 92, 09123 Cagliari, Italy\\
\vspace*{0.4cm}
INFN, Sezione di Cagliari\\
}
\end{center}
\vfill
\centerline{\bf Abstract}
\vskip 15mm

We investigate the gauging of a three-dimensional deformation of the
\ads algebra, which accounts for the existence of an invariant energy scale.
By means of the \psm formalism, we obtain explicit solutions of the field
equations, which reduce to the BTZ black hole in the undeformed limit.

\vfill
\end{titlepage}

\section{Introduction}

Theories of gravitation based on deformations of the \poi algebra
have been recently considered in two dimensions \cite{Mi1,Mi2} as a
possible implementation of ideas of deformed special relativity
(DSR) \cite{AC}-\cite{MS} to the domain of gravitational physics.
DSR aims at an effective description of the Planck scale physics
based on the hypothesis that the Planck energy is a fundamental
constant on the same footing as the speed of light, and
must therefore be left invariant under transformations of the frames
of reference. This implies that the \poi algebra is deformed at
microscopic scales \cite{AC}.

Deformed \poi algebras can be considered as special instances of
nonlinear algebras \cite{van}.
Although two-dimensional gravity theories based on nonlinear
algebras have been largely studied \cite{Ike}, the same cannot be
said about higher-dimensional models. This is because in more than
two dimensions one cannot define a natural action for the theory
without introducing auxiliary fields \cite{Iza}, whose physical
interpretation is not evident.

A useful tool for solving the field equations of nonlinear gauge
theories is the \psm formalism \cite{str}.
This has been employed till now only in the study of
\td gravity \cite{GKV}, but can be easily generalized for the
investigation of higher-dimensional models \cite{str2}.

In this paper, we consider an example of three-dimensional gravity
based on the deformation of the anti-de Sitter algebra introduced
in \cite{Mi2}, which generalizes the \dpa of \cite{MS} to the case
of nonvanishing cosmological constant. We discuss \bh and cosmological
solutions of the model by applying \psm techniques.

The \fe of nonlinear gauge theories are assumed to be of topological
type \cite{Ike}, since they require the vanishing of the field strength.
It is known that in three-dimensional riemannian geometry, the
condition of vanishing (constant) curvature is equivalent to
the Einstein equations with (non)vanishing cosmological constant
and hence, in the limit where the algebra is not deformed
we recover the known solutions of three-dimensional general
relativity. In this way we show the possibility of using the \psm
formalism to solve also higher-dimensional gravity.

The paper is organized as follows:
In sect.\ 2 we review the formalism of nonlinear gauge theories
and its application to gravity.
In sect.\ 3 we recover known results about 3D \ads gravity in the \psm
formalism. In sect.\ 4 we extend these results to deformed \ads
gravity.
A problem arises in the interpretation of the results due to the
lack of gauge invariance. We shall briefly discuss a possible
interpretation of this in the context of DSR, but a more detailed
discussion will be given elsewhere \cite{else}.
In sect.\ 5, we consider the limit $\l\to0$ of \poi gravity.

\section{Nonlinear gauge theories}

In this section, we review the definition of nonlinear gauge
theories and their application to gravity.

An algebra with generators $Q_A$ and commutation relations
\be
[Q_A,Q_B]=W_{AB}(Q),
\ee
where the structure functions $W_{AB}(Q)$ are regular functions of the generators,
antisymmetric in the two indices, that obey the generalized Jacobi identities
\be\lb{Jacobi}
{\de W_{[AB}\over\de Q_D}\,W_{C]D}=0,
\ee
is called nonlinear \cite{van}.

A gauge theory for this algebra \cite{Ike} can be defined by
introducing a gauge field $A^A$ and a coadjoint multiplet
of scalar fields $\F_A$, which
under infinitesimal \tran of parameter $\x^A$ transform as
\bea\lb{trans}
&&\d A^A=\di\x^A+U^A_{BC}(\F)A^B\x^C,\cr
&&\d\F_A=-W_{AB}(\F)\x^B,
\eea
where the $W_{AB}$ are now functions of the fields $\F^A$,
and the $U^A_{BC}$ are defined as
\bea
U^A_{BC}={\de W_{BC}\over\de\F_A}.
\eea

One can then define the covariant derivative of the scalar
multiplet
\be
\Di\F_A=\di\F_A+W_{AB}A^B,
\ee
and the curvature of the gauge fields
\be
F^A=\di A^A+\ha U^A_{BC}A^B\we A^C.
\ee
The variation of $F^A$ under a \gt is
\be
\d F^A=\left({\de W_{BC}\over\de\F_A}F^B-\ha{\de^2W_{BC}
\over\de\F_A\de\F_D}\,A^A\we\Di\F_D\right)\x^C.
\ee
Note that the gauge fields do not transform covariantly, due
to the second term. One can nevertheless define a covariant
derivative of the gauge fields as
\be
\Di F^A=\di F^A+{\de W_{BC}\over\de\F_A}A^B\we F^C-\ha{\de^2W_{BC}
\over\de\F_A\de\F_D}A^B\we A^C\we\Di\F_D,
\ee
This is easily checked to obey the Bianchi identity
\be
\Di F^A=0.
\ee

In order for the representation of the algebra to close, one must
impose the vanishing of the covariant derivative of the $\F$
fields,
\be\lb{fe1}
\Di\F_A=0.
\ee
Moreover, we require that the theory be of topological type, and
hence obey the \fe
\be\lb{fe2}
F^A=0.
\ee
In three dimensions, the \fe (\ref{fe1},\ref{fe2}) can be derived from
a BF-type lagrangian \cite{Iza},
 by introducing two auxiliary fields $C^A$ and $B_A$
\be
{\cal L}=\,^\ast\! C^A\we\Di\F_A+\,^\ast\! B_A\we F^A,
\ee
where $C^A$ and $B_A$ are 1- and 2-forms, respectively and the star denotes
the Hodge dual.

Deformed gravity on a three-dimensional manifold \man\ can be defined
starting from a six-dimensional nonlinear algebra, (e.g.
a deformation of the \trd \poi algebra), and identifying
three of the generators, which we shall denote as $P_\un{a}$,
$a=0,1,2$, with the generators of translations and
the other three, $M_\ov{a}$ with the generators of Lorentz
rotations.
One then identifies the components $A^\un{a}$ of the gauge fields with the
dreibeins $e^a$ and the components $A^\ov{a}$ with the spin connection $\o^a$.
The components of the gauge field strength can then be written in
terms of the geometric quantities of \man. We are mainly
interested in the special case in which the Lorentz group is
undeformed, while the momenta trasform nonlinearly, as in DSR
models. This implies
\bd
W_{\ov{a}\ov{b}}=\eabc\F_\ov{c},\quad W_{\un{a}\un{b}}=V_{ab}(\F_A),
\quad W_{\ov{a}\un{b}}=Y_{ab}(\F_A),
\ed
for arbitrary functions $V_{ab}$ and $Y_{ab}$.

The components of the gauge field strength are then
\be
F^\ov{a}=R^a+{\de V_{bc}\over\de\F_\ov{a}}\,e^b\we e^c+
{\de Y_{bc}\over\de\F_\ov{a}}\,\o^b\we e^c,
\ee
and
\be
F^\un{a}=T^a+{\de V_{bc}\over\de\F_\un{a}}\,e^b\we e^c+
\left({\de Y_{bc}\over\de\F_\un{a}}-\e^a_{\ bc}\right)\o^b\we e^c,
\ee
where $R^a=\di\o^a+\e^a_{\ bc}\,\o^b\we\o^c$ is the curvature and
$T^a=\di e^a+\e^a_{\ bc}\,\o^b\we e^c$ the torsion of \man.

The geometric quantities so defined in general are not
covariant under the full gauge group, even when $D\F_A=0$,
but only under a subgroup.
For example, in the case of undeformed \poi invariance, curvature
and torsion transform
covariantly only under the Lorentz subalgebra, while under
translations they transform one into each other.

As in ordinary gauge theories of gravity, one can also show that
\gct are equivalent to \gts on shell.
In fact, writing $A^A=A^A_\m\di x^\m$, under an infinitesimal
change of \coo of parameter $\z^\n$, the gauge fields transform
as standard vectors,
$\d_{C}A^A_\m=\de_\m\z^\n A^A_\n+\z^\n\de_\n A^A_\m$, while under
an infinitesimal \gt of parameter $\x^A$, $\d_GA^A_\m=\Di_\m\x^A$,
see (\ref{trans}).
Simple algebraic manipulations permit then to write $\d_C A^A_\m$ as
$\d_CA^A_\m=D_\m(\z^\n A^A_\n)+\z^\n F^A_{\n\m}$.
On shell, where $F^A_{\n\m}=0$, \gct are therefore equivalent to
\gts with parameter $\x^A=\z^\n A^A_\n$.

In order to solve the field equations, it is useful to adopt the
\psm formalism \cite{str}. Essentially, this formalism is based on
the identification of the fields $\F_A$ with the coordinates of a
Poisson manifold \cN with Poisson structure given by the functions
$W_{BC}(\F_A)$. The gauge
fields are then 1-forms on \cN and one can perform a change
of \coo on \cN to a Darboux basis where the \fe assume an almost trivial
form.

\section{Anti-de Sitter gravity}

Let us first consider the undeformed \ads algebra $so(2,4)$.
It satisfies
\bd
\{M_a,M_b\}=\eabc M_c,\qquad \{M_a,P_b\}=\eabc P_c,\qquad
\{P_a,P_b\}=\l^2\eabc M_c.
\ed
and admits two quadratic Casimir invariants:
\be
C_1=h^{ab}(P_aP_a+\l^2M_aM_a), \qquad C_2=\l\ h^{ab}M_aP_a,
\ee
where $h_{ab}=(-1,1,1)$ is the flat metric on \man.
It is convenient to define new generators
\be
\NN_a=M_a\pm\lm P_a,
\ee
which satisfy the $so(1,2)\per so(1,2)$ algebra
\be\lb{Nalgebra}
\{\NN_a,\NN_b\}=\eabc\NN_c, \qquad \{N^+_a,N^-_b\}=0.
\ee
In this basis the Casimir invariants are given by
\be
C_\pm=h^{ab}\NN_a\NN_b.
\ee
Likewise, one can define scalar fields
\be
\yy_a=\F_\ov{a}\pm\lm\F_\un{a}.
\ee

The choice of a Darboux basis is complicated by the existence of
different representations of $so(1,2)\per so(1,2)$, which give rise to
different solutions of the field equations.
The representations can be classified according to the sign of
$(\yy_1)^2-(\yy_0)^2$  and of $\yy_0$. Of course a great number of
subcases is possible, so we shall discuss only the most interesting.

\subsection{Black hole solutions: $(\yy_0)^2>(\yy_1)^2$.}

In the case $(\yy_0)^2>(\yy_1)^2$, $\yy_0>0$, a Darboux basis is  given
by
\be\lb{Dbasis}
\XX_1=h^{ab}\yy_a\yy_b,\qquad\XX_2=\yy_2,\qquad\XX_3=\arcch{\yy_0
\over\sqrt{(\yy_0)^2-(\yy_1)^2}}.
\ee
The new fields obey
\be\lb{Xcr}
\{\XX_1,\XX_2\}=\{\XX_1,\XX_3\}=0,\quad\{\XX_2,\XX_3\}=1, \quad
\{X^+_a,X^-_b\}=0.
\ee
The relations (\ref{Dbasis}) can be inverted, to obtain
\bd
\yy_0=\sqrt{(\XX_2)^2-\XX_1}\ \cosh\XX_3,\qquad
\yy_1=\sqrt{(\XX_2)^2-\XX_1}\ \sinh\XX_3.
\ed
In the basis (\ref{Dbasis}) the \fe take an elementary form
\cite{str} ($\a=1,2,3$):
\be\lb{Deq}
d\AA_\a=0,\quad\di\XX_1=0,\quad\di\XX_2=-\AA_3,\quad\di\XX_3=\AA_2,
\ee
and are solved by
\bea\lb{Dsol}
&&\XX_1=\const=\QQ, \qquad\XX_2=\yy,\qquad\XX_3=\hh,\cr
&&\AA_1=\ha\di\ww,\qquad\AA_2=\di\hh,\qquad\AA_3=-\di\yy,
\eea
where $\yy$, $\hh$ and $\ww$ are arbitrary functions.
Hence, $\yy_2=\yy$ and
\bd
\yy_0=\sqrt{(\yy)^2-\QQ}\cosh\hh,\qquad
\yy_1=\sqrt{(\yy)^2-\QQ}\sinh\hh.
\ed
Defining $\EE^a\id\ha\left(\o^a\pm\l e^a\right)$, one has
\bea
\EE^0&=&{\de\XX_\a\over\de\yy_0}\AA_\a=-2\yy_0\AA_1-
{\yy_1\over(\yy_0)^2-(\yy_1)^2}\ \AA_3,\cr
\EE^1&=&{\de\XX_\a\over\de\yy_1}\AA_\a=2\yy_1\AA_1+
{\yy_0\over(\yy_0)^2-(\yy_1)^2}\ \AA_3,\cr
\EE^2&=&{\de\XX_\a\over\de\yy_2}\AA_\a=2\yy_2\AA_1+\AA_2.\nn
\eea
One must now perform a gauge choice. Since the gauge algebra is
six-dimensional, one must fix three of the free functions in order
to obtain the \coo of a \trd spacetime.
Hence we set $\h^+=\h^-=0$; the third condition is imposed by
introducing a new variable $r$ such that $\yy=\mp\l r+{J\over 2r}$.
Moreover, we define $\QQ=M\mp\l J$.
It follows that
\be
\sqrt{(\y^+)^2-Q^+}=\sqrt{(\y^-)^2-Q^-}=\btz\id\G(r).
\ee
In this gauge the scalar fields read $\NN_0=\G$,
$\NN_1=0$, $\NN_2=\mp\l r+J/2r$, and then
\bd
\EE^0=-\G\di\ww,\quad
\EE^1=\left({J\over2r^2}\pm\l\right)\ {\di r\over\G},\quad
\EE^2=\left(\pm\l r+{J\over2r}\right)\di\ww.
\ed
It is easy to check that, if one puts $\ww=-(\f\pm\l t)$,
this coincides with the BTZ solution \cite{BTZ}. In fact, one has
\bea\lb{bh}
&&e^0=\G\di t,\quad e^1={\di r\over\G},\quad e^2=r\di\f-{J\over2r}\di t,\cr
&&\o^0=\G\di\f,\quad\o^1={J\over2r^2}{\di r\over\G},\quad\o^2=\l^2r\di t
-{J\over2r}\di\f.
\eea

As is well known, this solution corresponds to a spacetime of constant
curvature, with a conical singularity at the origin, and two horizons
at $r_\pm^2=(M\pm\sqrt{M^2-\l^2J^2})/2\l^2$. For $J=0$, it
reduces to the anti-de Sitter black hole with a single horizon at
$r_+^2=M/\l^2$.

Another interesting possibility occurs for $\y_0^+<0$, $\y_0^->0$.
In this case,
the definition of $\XX_3$ in (\ref{Dbasis}) is modified according to
\bd
\XX_3=\mp\arcch{\mp\yy_0\over\sqrt{(\yy_0)^2-(\yy_1)^2}}.
\ed
The \pb (\ref{Xcr}) are still satisfied. Going through the same steps
as before, one obtains $\yy_2=\yy$ and
\bd
\yy_0=\mp\sqrt{(\yy)^2-\QQ}\cosh\hh_3,\qquad
\yy_1=\mp\sqrt{(\yy)^2-\QQ}\sinh\hh_3.
\ed
Imposing the gauge conditions $\h^+=\h^-=0$,
$\yy=\l r\mp{J\over 2r}$ and defining $\QQ=M\mp\l J$,
one gets $\NN_0=\mp\G$,
$\NN_1=0$, $\NN_2=\l r\mp J/2r$, and \bd
\EE^0=\pm\G\di\ww,\quad
\EE^1=\left({J\over2r^2}\pm\l\right)\ {\di r\over\G},\quad
\EE^2=\left(\l r\mp{J\over2r}\right)\di\ww.
\ed
Putting $\ww=\l t\pm\f$, one recovers (\ref{bh}),
although in this case the scalar fields are different. This
will have important implications in the deformed case.

\subsection{Cosmological solutions: $(\yy_0)^2<(\yy_1)^2$.}

The solutions with $(\yy_0)^2<(\yy_1)^2$ can be interpreted either
as the region between the horizons of the black hole solutions of
the previous section, or as cosmological solutions.

They can be obtained in the same way as before, except that one
has to modify the definition of $\XX_3$ in (\ref{Dbasis}).
For example in the case $\yy_0>0$, one must define
\bd
\XX_3=\arcsh{\yy_0\over\sqrt{(\yy_1)^2-(\yy_0)^2}}.
\ed
A straightforward calculation yields
\bd
\yy_0=\sqrt{\QQ-(\yy)^2}\sinh\hh,\qquad
\yy_1=\sqrt{\QQ-(\yy)^2}\cosh\hh.
\ed
In the gauge $\hh=0$, $\yy=\mp\l r+{J\over 2r}$, the dreibein read
\be\lb{ibh}
e^0={\di r\over\G},\quad e^1=\G\di t,\quad e^2=r\di\f-{J\over2r}\di t.
\ee
The coordinate $t$ is now spacelike, while $r$ is timelike, and
the solution can then be interpreted as the interior of the \bh
(\ref{bh}).

However, a more interesting interpretation is as a cosmological
solution. In particular, taking for simplicity $\l=1$, $M=1$,
$J=0$ and defining $\t=\arccos\y$, $\ww=\f\pm\c$, one has
\bea\lb{cos}
&&e^0=\di\t,\quad e^1= \sin\t\ \di\c,\quad e^2=\cos\t\ \di\f,\cr
&&\o^0=0,\quad\o^1=\sin\t\ \di\f,\quad\o^2=\cos\t\ \di\c.
\eea
This is the anti-de Sitter cosmological solution in unusual coordinates.

\section{Deformed anti-de Sitter gravity}

We pass now to consider the deformed anti-de Sitter algebra
introduced in \cite{Mi2}. This is a generalization to the case of
nonvanishing \cco of the \dpa introduced in \cite{MS}.

We split the indices $a,b$ in one timelike, 0, and two
spacelike indices, $i,j=1,2$.
The Lorentz subalgebra is undeformed
\bd
\{M_a,M_b\}=\eabc M_c,
\ed
while
\bd
\{M_0,P_0\}=0,\quad \{P_i,P_j\}=-\l^2\pik^2\left(\e_{ij}M_0+
{\e_{ik}M_kP_j-\e_{jk}M_kP_i\over\k}\right),
\ed
\bd
\{M_i,P_j\}=-\left(\e_{ij}P_0-\e_{ik}{P_kP_j\over\k}\right),
\quad\{P_0,P_i\}=\l^2\pik^3\e_{ij}M_j,
\ed
\bd
\{M_i,P_0\}=\pik\e_{ij}P_j,\quad\{M_0,P_i\}=\e_{ij}P_j.
\ed
The algebra admits two quadratic Casimir invariants:
\bd
C_1=h^{ab}\left[{P_aP_a\over(1-P_0/\k)^2}+\l^2M_aM_a\right],
\qquad C_2={\l h^{ab}M_aP_a\over1-P_0/\k}.
\ed
Similarly to before, it is convenient to define new generators
\be
\NN_a=M_a\pm\lm\ {P_a\over1-P_0/\k},
\ee
which satisfy the algebra (\ref{Nalgebra}).

Likewise, one can define scalar fields
\be
\yy_a=\F_\ov{a}\pm\lm\ {\F_\un{a}\over1-\F_\un{0}/\k}.
\ee

Since the algebra satisfied by the $\yy_a$ is identical to that of
the previous section, one can proceed in the same way, but the relations
between with fields $\F_A$ are now more complicated:
\be
\F_\ov{a}=\y_a^++\y_a^-,\qquad
\F_\un{a}={\l(\y_a^+-\y_a^-)\over1+\l(\y_0^+-\y_0^-)/\k}.\quad
\ee

\subsection{Deformed black hole solutions.}

As we have seen, \bh solutions occur when $(\yy_0)^2>(\yy_1)^2$.
In this case, assuming $\yy_0>0$, one can use the Darboux \coo
(\ref{Dbasis}) obtaining
\bea
e^0&=&\sum_\pm{\de\XX_\a\over\de\F_\un{0}}\AA_\a={1\over\l}\sum_\pm
{\mp1\over(1-\F_\un{0}/\k)^2}\left(2\yy_0\AA_1+
{\yy_1\over(\yy_0)^2-(\yy_1)^2}\ \AA_3\right),\cr
e^1&=&\sum_\pm{\de\XX_\a\over\de\F_\un{1}}\AA_\a={1\over\l}\sum_\pm
{\pm1\over1-\F_\un{0}/\k}\left(2\yy_1\AA_1+
{\yy_0\over(\yy_0)^2-(\yy_1)^2}\ \AA_3\right),\cr
e^2&=&\sum_\pm{\de\XX_\a\over\de\F_\un{2}}\AA_\a={1\over\l}\sum_\pm
{\pm1\over1-\F_\un{0}/\k}\left(2\yy_2\AA_1+\AA_2\right),\nn
\eea
and for the connection,
\bea
\o^0&=&\sum_\pm{\de\XX_\a\over\de\F_\ov{0}}\AA_\a=-\sum_\pm\left(2\yy_0\AA_1+
{\yy_1\over(\yy_0)^2-(\yy_1)^2}\ \AA_3\right),\cr
\o^1&=&\sum_\pm{\de\XX_\a\over\de\F_\ov{1}}\AA_\a=\sum_\pm\left(2\yy_1\AA_1+
{\yy_0\over(\yy_0)^2-(\yy_1)^2}\ \AA_3\right),\cr
\o^2&=&\sum_\pm{\de\XX_\a\over\de\F_\ov{2}}\AA_\a=\sum_\pm\left(2\yy_2\AA_1+
\AA_2\right).\nn
\eea

Using the solutions (\ref{Dsol}) of the \fe and setting as before
$\h^+=\h^-=0$, one defines $\yy=\mp\l r+{J\over 2r}$,
$\ww=-(\f\pm\l t)$, $\QQ=M\mp\l J$. Substituting, one has
\bd
\F_\un{0}=\F_\un{1}=0,\quad\F_\un{2}=-\l^2r,
\ed
\bd
\F_\ov{0}=\G,\quad\F_\ov{1}=0,\quad\F_\ov{2}= J/2r.
\ed
It is then easy to see that the vielbein and the connection
are identical to (\ref{bh}). Hence, in this gauge the solution
are not deformed.

A less trivial situation occurs when $\y^+_0<0$, $\y^-_0>0$. Now,
setting $\h^+=\h^-=0$ and defining $\yy=\l r\mp{J\over 2r}$,
$\ww=\l t\pm\f$, $\QQ=M\mp\l J$, one has
\bd
\F_\un{0}=-\l\G/\D,\quad \F_\un{1}=0,\quad\F_\un{2}=-\l J/2r,
\ed
\bd
\F_\ov{0}=\F_\ov{1}=0,\quad\F_\ov{2}=\l r,
\ed
where $\D=1-\l\G/\k$ and $\G$ has been defined previously.
Moreover,
\be\lb{dbh}
e^0=\D^2\G\,\di t,\quad e^1={\D\over\G}\,\di r
\quad e^2=\D\left(r\,\di\f-{J\over2r}\,\di t\right),
\ee
and
\be
\o^0=\G\,\di\f,\quad\o^1={J\over2r^2}{\di r\over\G},\quad
\o^2=\l^2r\,\di t-{J\over2r}\,\di\f.
\ee
Hence, while the connection takes the same expression as in the
undeformed case, the vielbeins are deformed by factors of $\D$.
It follows that the solutions still have constant curvature, but
nontrivial torsion. In the limit $J=0$, for example, the torsion
reads
\be
T^0=-{3\l r\over\k\D^2}e^1\we e^0,\quad T^1=0,
\quad T^2=-{\l r\over\k\D^2}e^1\we e^2.
\ee

It is evident that the solutions have a singularity of the torsion
at $\D=0$, i.e.\
$r_+^2=\left(M+\k/\l+\sqrt{(M+\k/\l)^2-\l^2J^2}\right)/2\l^2$.
A coordinate singularity (horizon) is located at $r_-^2=(M+
\sqrt{M^2-\l^2J^2})/2\l^2$.
Since $r_+>r_-$ a naked singularity is always present.
These solution resemble the \td solutions of ref.\ \cite{Mi2}.

\subsection{Deformed cosmological solutions.}

The solutions with  $(\yy_0)^2>(\yy_1)^2$ can be discussed as in
the undeformed case. When $\yy_0>0$, in the gauge $\hh=0$ they take
again the form (\ref{ibh}) and can then be interpreted either as the
interior of the BTZ \bh or as a cosmological solution.

When $\y^+_0<0$, $\y^-_0>0$, the cosmological solution becomes more
involved. Proceeding as in section 2.2, one obtains for the
dreibein
\be
e^0=\D^2\di\t,\quad e^1=\D\sin\t\ \di x,\quad e^2=\D\cos\t\ \di\f,
\ee
where $\D=1-\sin\t/\k$, and again (\ref{cos}) for the connection.
The solution has constant curvature, but nontrivial torsion.
The nonvanishing torsion components in an orthonormal frame are
proportional to $1/\D^3$ and hence singular at $\sin\t=\k$. The solution
displays therefore a big bang singularity with the universe beginning
with a size of the order of the Planck length.

\subsection{Interpretation.}

Till now we have implicitly assumed that the properties of the
solutions of deformed gravity can be discussed in the gauge $\hh=0$.
However, as
has been noticed in sect.\ 2, the geometric quantities do not
transform covariantly under the full gauge group.
This implies for example that in general it is not possible to define a
spacetime metric invariant under all the gauge transformations\footnote
{A metric invariant under the action of the Lorentz group
can actually be constructed using also the
scalar multiplet $\F_A$, at the cost of introducing an additional
structure in the theory \cite{str2}.}.
In particular, in our case the standard metric is invariant under
rotations, but not under boosts.
This fact requires a physical interpretation compatible with the
postulates of DSR. Such interpretation must necessarily be different
from that of general relativity, and share some analogy with that of
Finsler geometry \cite{Asa}. A possible approach is to assume
that test particles
at rest in a given reference system experience a metric
$ds^2_0=h_{ab}e^ae^b$, where $h_{ab}$ is the usual flat metric,
while test particle moving with respect to the observer's frame
are seen to experience a different metric, given by the gauge
transformed one.
This topic will be discussed at length elsewere \cite{else}.

\section{The \poi limit}

The solutions for the gauge theory in the \poi
limit $\l=0$ of the anti-de Sitter algebra
can be obtained either from the previous results, taking the limit
$\l\to0$ after a suitable rescaling of the scalar fields $\yy$,
or repeating from the beginning the steps that led to the solution
in the \ads case.

Finding Darboux \coo for the \poi algebra is not trivial.
A possible choice is
\bd
X^+_1=\F_\un{a}\F^\un{a},\quad X^-_1=\F_\un{a}\F^\ov{a},\quad
X^+_2=\F_\un{2},\quad X^-_2=\F_\ov{2},
\ed
\bd
X^+_3={\F_\un{1}\F_\ov{0}-\F_\un{0}\F_\ov{1}\over\F_\un{1}^2-\F_\un{0}^2},
\qquad X^-_3=\arcch{\F_\un{0}\over\sqrt{\F_\un{1}^2-\F_\un{0}^2}},
\ed
which satisfy the algebra (\ref{Dbasis}). The \fe have therefore the form
(\ref{Deq}), with solutions (\ref{Dsol}). However, the relations between
the original fields and the $\XX_a$ are more involved:
\bd
\F_\un{0}=\g\cosh X^-_3,\quad\F_\ov{0}=\g X^+_3\cosh X^-_3
+\g^\mo(X^-_1-X^+_2X^-_2)\sinh X^-_3,
\ed
\bd
\F_\un{1}=\g\sinh X^-_3,\quad\F_\ov{1}=\g X^+_3\sinh X^-_3
+\g^\mo(X^-_1-X^+_2X^-_2)\cosh X^-_3,
\ed
where $\g=\sqrt{X^+_1-(X^+_2)^2}$.

In the gauge $\hh=0$, $\y^+\y^-=J/2$, setting $Q^+=M$, $Q^-=J$,
$\y^+=r$, $\o^+=t$, $\o^-=\f$,
one recovers solutions of the form (\ref{bh}), but with
\be\lb{flat}
\G(r)=\sqrt{M+{J\over2r^2}}.
\ee
These are the solution of ref.\ \cite {J} in different coordinates,
and describe flat spacetime with a conical singularity at the
origin.

In the deformed case, for $\l\to0$, the deformed algebra reduces to the
\dpa of ref.\ \cite{MS}.
Again, the solutions can have the form (\ref{bh}) or (\ref{dbh}), but
with $\G$ given by (\ref{flat}).

We also remark that solutions in the case of positive \cco can be
obtained by analytic continuation $\l\to i\l$ of the \ads
solutions, but we shall not discuss them here.

\section{Conclusions}

We have shown how to obtain the solutions of deformed
three-dimensional models of gravity by using the \psm formalism.
In the case of \ads or \poi invariance we recover the results
obtained by standard methods.
The discussion of the physical properties in the deformed case
is not straightforward, since the geometry does not display the
same invariance as the gauge theory, and depends on an interpretation
of the model, which will be discussed in detail elsewhere \cite{else}.

The same techniques could be applied to more standard
deformations of the gauge algebra that preserve Lorentz invariance,
analogous to the two-dimensional models studied in \cite{GKV}.
A drawback of this formalism is however that there is no
natural way to fix the gauge conditions on the fields. We have
chosen them in such a way to obtain the known results in the
undeformed limit, but it is difficult to give a general rule
that clarifies the physical meaning of specific gauge choices
and their relation with coordinate transformations.

\bigskip
\noindent{\bf Acknowledgments}
\smallskip

I wish to thank T. Strobl for useful discussions. This work was
partially supported by ESI.


\begin{thebibliography}{99}
\bibitem{Mi1}
S. Mignemi, \PL{B551}, 169 (2003).
\bibitem{Mi2}
S. Mignemi, \MPL{A18}, 643 (2003).
\bibitem{AC}
G. Amelino-Camelia, \IJMP{D11}, 35 (2002).
\bibitem{dpa}
J. Lukierski, A. Nowicki, H. Ruegg and V.N. Tolstoy, \PL{B264},
331 (1991); S. Majid and H. Ruegg, \PL{B334}, 348 (1994).
\bibitem {MS}
J. Magueijo and L. Smolin, \PRL{88}, 190403 (2002).
\bibitem{van}
K. Schoutens, A. Sevrin and P. van Nieuwenhuizen, \IJMP{A6}, 2891
(1991).
\bibitem{Ike}
N. Ikeda, \AoP{235}, 435 (1994).
\bibitem{Iza}
K.-I. Izawa, \PTP{105}, 225 (2000).
\bibitem{str}
P. Schaller and T. Strobl, \MPL{A9}, 3129 (1994);
T. Kl\"osch and T. Strobl, \CQG{13}, 965 (1996).
\bibitem{GKV}
D. Grumiller, W. Kummer and D.V. Vassilevich,
\PRep{369}, 327 (2002).
\bibitem{str2}
T. Strobl, \hep{0310168}.
\bibitem{else}
S. Mignemi, in preparation.
\bibitem{BTZ}
M. Ba\~nados, C. Teitelboim and J. Zanelli, \PRL{69}, 1849 (1992).
\bibitem{J}
S. Deser, R. Jackiw and G. 't Hooft, \AoP{152}, 220 (1984).
\bibitem{Asa}
G.S. Asanov, {\it Finsler geometry, relativity and gauge
theories}, D. Reidel 1985.

\end{thebibliography}
\end{document}